\documentclass[12pt]{amsart} 
\usepackage{geometry}
\usepackage{bm}
\usepackage{graphicx}
\usepackage{graphics}
\usepackage{float}
\usepackage{amsmath}
\usepackage{amssymb}
\usepackage{amsfonts}
\usepackage{latexsym}
\usepackage{epsfig}
\usepackage{dcolumn}
\usepackage{url}

\usepackage[francais,english]{babel}
\usepackage{lmodern} 
\usepackage[T1]{fontenc}
\usepackage{pdfsync}
\usepackage[pdftex,bookmarks,colorlinks,breaklinks]{hyperref}
\hypersetup{linkcolor=black,
citecolor=blue,filecolor=red}

\newcommand{\Ecal}{{\mathcal E}}

\newcommand{\Ical}{{\mathcal I}}
\newcommand{\Hcal}{{\mathcal H}}

\newcommand{\Tcal}{{\mathcal T}}

\newcommand{\half}{{\textstyle \frac{1}{2}}}

\newcommand{\rvec}{{\textbf r}}
\newcommand{\svec}{{\textbf s}}

\newcommand{\nvec}{{\textbf n}}

\newcommand{\grad}{{\boldsymbol \nabla}}

\newcommand{\Avec}{{\textbf A}}
\newcommand{\Bvec}{{\textbf B}}

\newcommand{\rot}{{\nabla \times}}

\usepackage{epstopdf}

\title{\textsf{\textbf{Flux ropes as singularities of the vector potential}}}

 \author{Maurice Kleman}
\begin{document}
\maketitle
\begin{center}
Institut de Physique du Globe de Paris, Sorbonne Paris Cit\'e, \\
1 rue Jussieu, Paris c\'edex 05, France \footnote{e-mail: kleman@ipgp.fr}
\end{center}
\scriptsize

\scriptsize\begin{abstract}
A flux rope is a domain of concentration of the magnetic field $\textbf{B}$. Insofar as $\textbf{B}$ outside such a domain is considered as vanishingly small, a flux rope can be described as the core of a singularity of the outer vector potential $\textbf{A}$, whose topological invariant is the magnetic flux through the rope. By 'topological' it is meant that $\oint_C\textbf{A}\cdot\mathrm d\textbf{s}$ measures along any loop $C$ surrounding the flux rope the same constant flux $\Phi$. The electric current intensity is another invariant of the theory, but non-topological. We show that, in this theoretical framework, the linear force-free field (LFFF) Lundquist model and the non-linear (NLFFF) Gold$-$Hoyle model of a flux rope exhibit stable solutions distributed over quantized strata of increasing energies (an infinite number of strata in the first case, only one stratum in the second case); each stratum is made of a continuous set of stable states.   
The lowest LFFF stratum and the unique NLFFF stratum come numerically close one to the other, and match with a reasonable accuracy the data collected by spacecrafts travelling across magnetic clouds. The other LFFF strata do not match these data at all. It is not possible at this stage to claim which model fits better the magnetic cloud data. We also analyze in some detail the merging of tubes belonging to the same stratum, with conservation of the magnetic helicity, and the transition of a tube from one stratum to another one, which does not conserve magnetic helicity.
\end{abstract}
 \newpage\normalsize

\section{\sf {\scriptsize INTRODUCTION}}\label{s1}
It has been shown in \cite{kleman14} that a {\it  {singular irrotational} magnetic field} can behave as a source for twisted flux tubes (flux ropes). The singularity we have in mind carries an electric current  $\Ical =(c/4\pi)\oint_C\Bvec\cdot\mathrm d\svec$, where $C$ is any loop surrounding the flux rope.
 A full analysis of such an electromagnetic configuration relies on the relationship between the topology of a three-dimensional domain $\Omega$ and the calculus of divergenceless vector fields $\Bvec$ defined in such a domain, under the condition that $\Bvec_n =0$ on $\partial\Omega$ (tangential boundary conditions); these {\it harmonic knots} (so qualified because $\Delta \Bvec=0$, except on the singularities) are described in mathematical terms in \cite{cantarella02} (where more general vector fields are also described). How the theory applies to physical situations (magnetic fields in space, in the shape of flux tubes) is summarized in Sect.~1.1; the notions to be developed in this article are introduced at the end of this section, para. c) and d). Sect.~1.2 summarizes the content and main results developed in this article.\bigskip

1.1$-$ {\sf Main concepts in the singularity theory of flux ropes}\smallskip

a)$-$ \textsf{The current intensity as the topological invariant}: In this article we shall restrict our considerations to the most simple harmonic knot, namely a domain $\Omega$ made of the whole 3D space pierced by an infinite cylindrical hole $\Tcal$ along the $\bf z$-axis.
According to the fundamental properties of harmonic knots \cite{kleman14,cantarella02}, the irrotational field $\Bvec$ is uniquely determined by the following data

i$-$ boundary conditions
\begin{equation}
\label{e11}
\Bvec_n = 0 \ \mathrm{on}\ \partial \Tcal\qquad\Bvec = \{0, 0,B_o\} \ \mathrm{at \ infinity}
\end{equation}

We shall use cylindrical coordinates all along, such that $\Bvec=\{B_r, B_\theta,B_z\}$.

ii$-$ a line integral $\beta$ along a closed curve $C$ surrounding $\Tcal$ \cite{kleman14,mahajan98}
\begin{equation}
\label{e1}
\beta = \oint_{C}\Bvec\cdot\mathrm{d}\svec.
\end{equation} 
 By Stokes' theorem, the quantity $\beta$ takes the same value if $C$ is replaced by a homologous loop $C'$ ($\Bvec$ harmonic implies $\rot\Bvec=0$): we call $\beta$ a {\sf topological invariant}. In the limit where the radius of $\Tcal$ is reduced to a line $L$ (which we choose for simplicity to be
along the $\bf z$-axis), $\Bvec$ becomes singular on $L$, and we may write
\begin{equation}
\label{e2}
\rot\Bvec= \beta\bf{\hat{z}}\delta^2_{\textit{L}}(\rvec).
\end{equation}
$\Bvec$ can be expressed as the gradient of a multivalued scalar field $\psi$, $\Bvec=\grad\psi$
\begin{equation}
\label{e44}
\psi= \frac{\beta}{2\pi}\theta + \gamma z 
       \qquad \Bvec= \{
  0,     \frac{ \beta}{2\pi  r}, \gamma\},
\end{equation}
where $\beta$ is related to the electric current along $L$: $\beta=({4\pi}/{c})\Ical$ cf. Eq.~\eqref{e1}, $\gamma = B_o$. The level sets $\psi=constant$ are ruled helicoids of pitch $-b$, with\footnote{a right helix carries a positive pitch, a left helix a negative pitch, by convention, which choice implies the use of a right-handed coordinate frame.} \begin{equation}
\label{e3}
b= \frac{\beta}{\gamma}=\frac{4\pi}{c}\frac{\Ical}{B_o}.
\end{equation} Thereby one can write
\begin{equation}
\label{e4}
 \Bvec= \{
  0,     \frac{ B_o b}{2\pi  r}, B_o\}
\end{equation}
\indent These properties are {reminiscent of the theory of {\it defects} in condensed matter physics} \cite{klfr08}; $L$ has the status of a vortex line endowed with a helical structure $-$ which character makes it also akin to a screw dislocation line \cite{kleman14} $-$ a typical defect defined by a scalar invariant, here $\Ical$. In the example above, $L$ is an infinite straight line, but it can be any closed loop or any curve going both ends to infinity, that carry a {topological invariant} defined by a line integral as Eq.~\ref{e1}, taken along any loop $C$ surrounding the singularity and not enclosing another singularity
\begin{equation}
\label{e77}
\frac{4\pi}{c}\Ical = \oint_C\Bvec\cdot\mathrm d\svec.
\end{equation}

b)$-$ \textsf{The core:} 
The foregoing captures the essential topological properties of an harmonic knot, but it is of course an impossible physical object, due to the presence of the $1/r$ behaviour in \eqref{e4} (equivalently, the presence of a delta function in \eqref{e2} tells us about this impossibility). This can be cured by drilling a cylindrical hole along $L$, thus coming back to $\Omega$; the singularity, which obeys \eqref{e4}, is then virtual in the vacuum of the hole. Another possibility is to fill the hole with a magnetised plasma carrying an electric current $(c/4\pi)\beta$, $-$ we shall call this region the {\sf core} of the singularity $-$ such that some boundary conditions (to be discussed soon) are satisfied at the contact between this {core} $\Tcal$ and the outside medium where $\Bvec_o$ obeys Eq~\ref{e4} (from now on, we shall use $\Bvec_o$ to denote the field outside the core, $\Bvec_i$ the field inside the core). 

This configuration provides a model for a \textit{twisted tube} (a flux rope), because the boundary conditions, whatever they may be, necessarily transmit to the core the helical nature of $\Bvec_o$. This can take on at least two aspects: either the magnetic field $\Bvec$ is continuous across $\partial\Tcal$, or else the magnetic field is allowed to be discontinuous (which requires the presence of a surface current for $r = r_0$). In both cases the vector potential $\Avec$ is continuous, cf \cite{griffiths99} (a discontinuity of the vector potential would imply a $\delta$-function singularity in the $\Bvec$-field on the boundary (see the discussion in \cite{kleman14}, footnote 2), which is most probably not physical, and will not be considered.)\smallskip

c)$-$ {\sf Quantized strata}:  
 There are {\it restrictions} on the allowed field configurations; they have indeed to minimize the magnetostatic energy $\Ecal = \frac{1}{8\pi}\iiint_\Omega (B_\theta^2 +B_z^2)\mathrm dV$ $-$ under the assumption that the field 'carries' a topological invariant ($\Ical$ in the case above, Eq.~\eqref{e77}) $-$  and that the boundary conditions (continuity of $\Bvec$ for $r= r_0$ in the first case above; at infinity $\Bvec= \{0,0,B_o\}$) are obeyed. The calculations yield 
critical points satisfying 

i$-$ $\partial \Ecal / \partial r_0=0$ (the condition of stability $\partial^2\Ecal/\partial r_0^2 >0$ has to be checked) 

 ii$-$ ${B_{i,\theta}}/{B_{i,z}}(r_0) = {B_{o,\theta}}/{B_{o,z}}(r_0)= {b}/{2\pi r_0}$ (boundary conditions).
 
  One gets, for a given current intensity $\Ical$, separated sets of solutions, {\sf quantized strata} so to speak: the tube radii, the pitches, the helicities and the energies are distributed continuously over each of these sets. See \cite{kleman14} and what follows for examples.

 The quantization rules depend on the core model, {\it e.g.}, linear force-free field, uniformly twisted field (non-linear), constant current density, linear azimuthal current, $\dots$ \cite{dasso05}. Here, we investigate the first two models. \smallskip

   d)$-$ {\sf The flux as the topological invariant, the AB model}: We have dwelt on the case where the topological invariant is the current; this requires that  the outside field be irrotational, but non-vanishing. The assumption of irrotationality is often made, but even more often is it assumed in the numerical simulations that $\Bvec_o$ totally vanishes, which makes simulations easier and is justified by the extremely small field measured outside the ropes, at the limit of measurement possibilities \cite{stenflo78}. Recent measurements made by spacecrafts outside flux ropes 
   look like fluctuations 
 or perhaps flux ropes of small radius and small flux: the field obviously concentrates in the ropes themselves. Therefore it is tempting to consider that the relevant variable is no longer the magnetic field $\Bvec$, but the {\sf vector potential} $\Avec, \ \Bvec = \rot\Avec$, irrotational outside the core. Aharonov and Bohm \cite{aharonov59} have recognized that this configuration yields interesting quantum effects involving interference patterns between electron currents passing next to a (small $-$ in their case) solenoid; we do not expect to observe such effects in space, whose evidence would unambiguously prove, if they exist, that a space flux rope is a physical realization of a AB (Aharonov$-$Bohm) model. 
      
   In the Coulomb gauge $\grad\cdot\Avec =0$ one has $\Avec_{o,n} =\Avec_{i,n} =0$, because of the continuity of the vector potential at a boundary \cite{griffiths99}. Inside the tube the magnetic field $\Bvec_{i}$ is continuous, outside $\Bvec_{o}=0$; the magnetic field discontinuity at the boundary requires a surface current that compensates for the volume current inside; the total current has to vanish \cite{kleman14}. With such a picture the relevant \textit{topological invariant} is no longer the intensity $\Ical$ but the flux $\Phi$.
\begin{equation}
\label{e5}
\Phi=\oint_C\Avec\cdot\mathrm d\svec.
\end{equation}   

The concept of core is of course relevant here as in the previous case. 

We shall see that in the minimization process the inner total current intensity (equal $-$ and opposite in sign $-$ to the current flowing along the boundary) still plays a fundamental role. This is discussed in Sect.~\ref{s12} in connection with the case when the outside magnetic field is small compared to the inside magnetic field and tends to zero smoothly, {\it i.e.,} a configuration where the topological invariant $-$ the total current intensity as long as $\Bvec_o\neq 0$ $-$  turns suddenly to the other topological invariant, the flux, when $\Bvec_o=0$. At the same time the field configuration changes \textsf{continuously}. This is a further justification of the AB approach: \textsf{the domain outside a flux rope}, little magnetized as it is, \textsf{can be treated as a perturbation of the AB model, i.e., $\Avec_o$ irrotational}. 
\bigskip

1.2$-${\sf Content of this article}\smallskip

As in \cite{kleman14} we require the Lorentz force to vanish at any point of the core; such a force-free field obeys the Beltrami equation
\begin{equation}
\label{e99}
\ell \rot \Bvec = \Bvec, 
\end{equation} where $\ell$ is a constant on each line of force of the magnetic field, since $\grad\cdot\Bvec=0$. The flux rope will be supposed to be a circular cylinder with translational symmetry along its axis. 

We explore two situations where the topological invariant is the {\it flux}, {\it i.e.,} the vector potential $\Avec_o$ is irrotational, $\Bvec_o =0$:\smallskip

 i$-$ the Lundquist {\it linear} force-free field case (LFFF) \cite{lundquist50} $-$ the linearity implies that $\ell$ is a constant over all the rope $-$. In Sect.~\ref{s12} we show that the Bohm$-$Aharonov model is the continuous limit of a model where $\Bvec$ is discontinuous for $r=r_0$, when $B_o\rightarrow 0$ Sect.~\ref{s2}. Thus the AB model is yet an excellent approximation of a true flux rope. 

 ii$-$ the Gold$-$Hoyle non-linear force-free field case (NLFFF) \cite{gold60}, Sect.~\ref{s3}, where $\ell$ depends on radius.\smallskip
  
  In the same two sections the processes of tube merging and tube transitions are analyzed. Flux ropes belonging to some stratum tend to merge while keeping in the same stratum and thus release large amounts of energy; a rope belonging to a higher stratum release some energy when falling to the lowest stratum. These two processes are very different: the merging of a collection of ropes proceeds quasi continuously at constant magnetic helicity; the transition of a rope from a stratum to another involves topological modifications which do not conserve magnetic helicity.\smallskip

  In Sect.~\ref{s4} we compare some observational data for {\sf magnetic clouds} \cite{lepping97,dasso03,demoulin14,dasso05} to the theoretical results developed in Sect.~\ref{s12}, \ref{s2} and \ref{s3}. {Magnetic clouds (MC)} are flux ropes that extend between the Sun and the Earth, and which are triggered by the corona mass ejections (CME) that originate at the surface of the Sun. It turns out that the lowest stratum, in the LFFF case, seems to match the observational results with a good accuracy, whereas all the higher strata do not. In the NLFFF case there is only one stratum, which is very siilar to the LFFF lowest stratum. Thus it appears difficult to decide which model suits better the observations, but it is already a remarkable fact that the observations are so well fitted, and that no higher stratum than the lowest one is observed, a result that also transpires from the theory. 

\section{\sf {\scriptsize THE AB MODEL AS A CONTINUOUS LIMIT OF THE $\Bvec-$DISCONTINUOUS MODEL.}}\label{s12}
In this section, it is assumed that $B_i$ and $B_o$ are discontinuous on the boundary $r=r_0$, cf. \eqref{e4} for the expression of $\Bvec_o$. The topological invariant is the total current intensity $(c/4\pi)B_o\ b =(c/4\pi)\oint_C\Bvec\cdot\mathrm d\svec$. The limit configuration $B_o=0$ is the same as that we discuss in Sect.~\ref{s2}.
 We use the LFFF model; $\Avec$ is written in the Coulomb gauge. \bigskip

2.1$-$ \textsf{magnetic field and vector potential}\smallskip

 a)$-$ \textsf{outside:} 
\begin{equation}
\label{e10}
\Bvec_o =\{0,\frac{B_ob}{2\pi r}, B_o\},\qquad \Avec_o =\{0,\frac{B_or}{2} + \frac{c_\theta}{r}, -\frac{B_ob}{2\pi }\ln{\frac{r}{r_0}}+d_z\}
\end{equation}
where $c_\theta$ and $d_z$ are two constants determined below by the boundary conditions at $r=r_0$; these constants enter in the components of the irrotational part of $\Avec_o$,\smallskip

b)$-$ \textsf{inside:} \begin{equation}
\label{ }
\Bvec_i =\{0,AJ_1(\frac{r}{\ell}), AJ_0(\frac{r}{\ell})\},\qquad \Avec_i =\ell\{0,AJ_1(\frac{r}{\ell}), AJ_0(\frac{r}{\ell})\},
\end{equation}
where $\ell$ is the length associated to the Beltrami equation.\bigskip

2.2$-$ \textsf{boundary conditions}\smallskip 

The magnetic field is discontinuous, but the vector potential has to be continuous, as emphasized previously Sect.~1.1b; thus 
\begin{equation}
\label{e47b}
c_\theta = -\half r_0^2 B_o +r_0 \ell AJ_1(\frac{r_0}{\ell}), \qquad  d_z = \ell AJ_0(\frac{r_0}{\ell})\end{equation}

The flux inside the tube $\Phi_i=\oint_{r=r_0}\Avec\cdot\mathrm d\svec $ can be calculated either with $\Avec = \Avec _i$ or $\Avec = \Avec _o$. One gets
$$\Phi_i  = 2\pi r_0 A J_{1} ( \frac{r_0}{\ell} ) = B_o \pi r_0^2+2\pi c_\theta .$$

Now comes an important property of the present model, namely that the pitch of the $\Bvec_i$-lines at the boundary $r=r_0$ must be the same as that of the $\Bvec_o$-lines, \textit{i.e.}, $\Bvec_o(r_0) = \lambda \Bvec_i(r_0)$, where $\lambda$ is some constant. If it were otherwise, the inner and outer configurations could be chosen independently one from the other, and this is precisely what we do not wish. This condition reads 
\begin{equation}
\label{e47c}
\frac{B_{o,\theta}(r_0)}{B_{o,z}(r_0)}=\frac{B_{i,\theta}(r_0)}{B_{i,z}(r_0)}=\frac{b}{2\pi r_0}.
\end{equation}  

Thereby let us introduce the two dimensionless parameters $\eta$ and $\zeta$ and a magnetic field $B_i$, all three characteristic of the flux rope $-$ these parameters will soon appear as the fundamental parameters of the problem:
 \begin{equation}
\label{e11c}
\eta = \frac{r_0}{\ell},  \ \zeta= \frac{J_1(\eta)}{J_0(\eta)}, \  B_i = AJ_0( \eta),
\end{equation}
With these notations $\Phi_i$ can be written
$$\Phi_i =2\pi r_0 \ell B_i \zeta. $$ 

Eventually, after further use of the boundary conditions \eqref{e47b} and \eqref{e47c}:
\begin{equation}
\label{ }
{\Phi_i = B_i b \ell,\qquad d_z = B_i\ell,\qquad 2\pi c_\theta = B_i b \ell -B_o \pi r_0^2}, \qquad \zeta=\frac{b}{2\pi r_0}.
\end{equation}
\smallskip

2.3$-$ \textsf{{electric currents, energies}}\smallskip

The inside current intensity 
\begin{equation}
\label{ }
\Ical _i= \frac{c}{4\pi}B_ib
\end{equation} is a constant.
  The total current, including the current flowing along the boundary $\Ical_s=(c/4\pi)(B_o-B_i)b$, due to the $\Bvec$-discontinuity at $r=r_0$, is 
 \begin{equation}
\label{ }
\Ical=\frac{c}{4\pi}B_ob.
\end{equation} 
 This total current is the topological invariant of the model, {i.e.,} $B_ob=\oint \Bvec\cdot\mathrm d \svec$, when $\Bvec$ traverses any loop surrounding the flux rope; $B_o$ is a given parameter (it will be made to vanish at constant $B_i$), thus $b$ is also given, as well as $B_i$. When $B_o=0$, the topological invariant vanishes, but another topological invariant appears, namely the magnetic flux $\Phi_i = \oint \Avec\cdot\mathrm d \svec =B_ib\ell$, where $\Avec$ traverses any loop surrounding the flux rope. \smallskip

The outside $\Ecal_o$ and inside $\Ecal_i$ energies can be written:
\begin{equation}
\label{eq41}
 \Ecal_{o} = \frac{B_o^2}{4}\left(\half(R^2-r_0^2)+(\frac{b}{2\pi})^2\ln{\frac{R}{r_0}}\right),\qquad \Ecal_{i} = \frac{B_i^2}{4}\left(r_0^2+(\frac{b}{2\pi})^2 -\frac{b\ell}{2\pi}\right),
\end{equation} where ${ R }$ is the outer radius.
 \vspace{2pt}

2.4$-$ \textsf{critical points} \smallskip

Let $\Ecal = \Ecal_{o} +\Ecal_{i}$. The derivative $\partial \Ecal/\partial r_0 = 0$ can be written:
\begin{equation}
\label{eq43}
\frac{\partial \Ecal}{\partial r_0}= \frac{r_0}{4}\left(   (2B_i^2 -B_o^2)+B_i^2 \zeta \frac{\zeta+\zeta^{-1}}{D}-B_o^2 \zeta^2\right)=0,
\end{equation} $\mathrm{where}\  D = 1-\eta(\zeta + \zeta^{-1}),$ $-$ we use the same notation as in \cite{kleman14}.
Assuming $D\neq 0$, one gets:
\begin{equation}
\label{eq44}
 (2B_i^2 -B_o^2)+B_i^2 (1+ \zeta^2)-B_o^2 \zeta^2=\eta(\zeta+\zeta^{-1}) (2B_i^2 -B_o^2 -B_o^2 \zeta^2)=0.
\end{equation} 

There are two extreme cases, and intermediary ones

i$-$ continuity of $\Bvec$; {$|B_o|= |B_i|$}. One gets $\eta = 2\zeta/(1-\zeta^4)$, a situation which is completely treated in \cite{kleman14}.\smallskip
 
ii$-$ $\Avec$ irrotational; {$B_o= 0$}. One gets $\eta = \zeta/2 +\zeta/(1+\zeta^2)$, which is the same expression as that in Sect.~\ref{s2}, Eq. \eqref{e37b}, also treated in \cite{kleman14}.\smallskip
   
iii$-$ In between, the configurations close to $B_o=0$ are the most interesting. Let us write
   $$B_o^2 =\epsilon B_i^2,$$ where $\epsilon >0$ is a small parameter.
Eq.~\eqref{eq44} becomes:
\begin{equation}
\label{eq45}
 3+ \zeta^2-\epsilon(1+ \zeta^2)=2\eta(\zeta+\zeta^{-1}) \left(1 -\half\epsilon(1+ \zeta^2)\right).
\end{equation} We write \begin{center}
$ \left(1 -\half\epsilon(1+ \zeta^2)\right)^{-1} \approx \left(1+ \half\epsilon(1+ \zeta^2)\right)$,

$\left(3+ \zeta^2-\epsilon(1+ \zeta^2)\right) \left(1+ \half\epsilon(1+ \zeta^2)\right) \approx 3+ \zeta^2+\half \epsilon(1+ \zeta^2)^2.$

\end{center} One eventually gets:
\begin{equation}
\label{eq46}
\eta=\frac{\zeta}{2} +\frac{\zeta}{1+\zeta^2}+\frac{\epsilon}{4}\zeta(1+ \zeta^2).
\end{equation} 

The configurations are continuously varying when $B_o \rightarrow 0$. Thus the limit case $\epsilon =0$ $-$ when the vector potential is irrotational and the fundamental invariant is the magnetic flux $-$ describes but to a small difference a configuration $\epsilon\neq0$ with a small magnetic flux outside the flux rope. \textsf{Thereby a $\epsilon\neq0$ field configuration can be studied as a perturbation of the limit configuration $\epsilon =0$, for which $\Avec_o$ is irrotational.}

\section{\sf {\scriptsize LINEAR FORCE-FREE FIELD IN THE AB FRAMEWORK }}\label{s2}

3.1$-$ \textsf{Magnetic field and vector potential; stable critical points}  \smallskip


    We use the results of the previous subsection with $B_o=0$. $\Avec_o$, which is obtained from \eqref{e10}, can now be written
$$\Avec_o= \Phi ( 0,\  \frac{1}{2\pi r}, \ \frac{1}{b}),$$ and satisfies the topological relationship
$$\Phi=\oint_C\Avec_o\cdot\mathrm d\svec.$$Thus the topological invariant is $\Phi$. 
In the previous section, the topological invariant $\Ical$ was split into two topological invariants $\Ical_i$ and $\Ical_s$, both constant in the minimization process. We conserve this invariance by reason of continuity, when $B_o\rightarrow 0$. But remember that the current intensity is no longer a topological invariant, it is however a constant in the energy minimization.

    The results are the same as in \cite{kleman14}, Sect. 5, but we present them in a simpler form. Also, we shall expatiate in the next subsection on the phenomena of tube merging and transitions between tube states. 
    
    We drop the index $i$ of $\Bvec_i$ and $\Phi_i=B_ib\ell$, as it is no longer necessary.\smallskip

The energy density is 
\begin{equation}
\label{e22}
\Ecal = \frac{B^2b^2}{16\pi^2}(1+\frac{1}{\zeta^2}-\frac{2\pi \ell}{b})= \frac{\Phi^2}{16\pi^2}(\frac{1}{\ell^2}+\frac{1}{\zeta^{2}\ell^2}-\frac{2\pi }{\ell b}),
\end{equation}which can also be written
 \begin{equation}
\label{e16}
{\Ecal = -\frac{B\Phi}{8\pi}D,} 
\end{equation}    
 because  of the identities
$({1}/{\ell^2})\Big\{1+{1}/{\zeta^2}-{2\pi \ell}/{b}\Big\}=({1}/{\ell^2 \eta \zeta})\Big\{-1+\eta(\zeta + \zeta^{-1} )\Big\}$
and $\ell^2 \eta \zeta = {\Phi}/({2\pi B}).$

The minimization of the free energy \eqref{e16} with respect to $r_0$ requires that some parameter, {\it e.g.}, $B$, $b$, or $\ell$ be fixed, besides the topological invariant $\Phi$. As explained above, it will be $\Ical_i \propto Bb$. Equivalently, it can be $\ell$, since $\Phi= Bb\ell$ is a constant. It is therefore equivalent to minimize with respect to $\eta = r_0/\ell$ or with respect to $r_0$. 
 To do so, we need the derivative
\begin{equation}
\label{e18a}
\frac{\mathrm{d}\zeta}{\mathrm{d}\eta} =1-\frac{\zeta}{\eta}+\zeta^2.
\end{equation} 
which can be derived as follows. We have 
$$\frac{\mathrm{d}\zeta}{\mathrm{d}\eta}= \frac{\mathrm{d}}{\mathrm{d}\eta}(\frac{J_1(\eta)}{J_0(\eta)})=\frac{J'_1(\eta)}{J_0(\eta)}-\frac{J_1(\eta)J'_0(\eta)}{J^2_0(\eta)} =\frac{J_0(\eta)-\frac{1}{\eta}J_1(\eta)}{J_0(\eta)}+\frac{J_1^2(\eta)}{J^2_0(\eta)},$$ where we have used \cite{abram} for the expressions of the derivatives of the Bessel functions. Then replace ${J_1(\eta)}/{J_0(\eta)}$ by $\zeta$ in this equation.

 The free energy can now be written
 $$\Ecal_\ell = \frac{\Phi^2}{8\pi \ell}(-\frac{D}{b}),$$ and the minimization  yields\footnote{We shall denote any quantity $f$ evaluated at a critical point as $f|_{r_0}$. For example $\frac{\partial D}{\partial \eta}|_{r_0}=0$.}

\begin{equation}
\label{e36b}
\frac{\mathrm{d}\Ecal_\ell}{\mathrm{d}\eta}|_{r_0} = 0=\frac{\Phi^2}{8\pi \ell}\frac{\mathrm{d}}{\mathrm{d}\eta}(-\frac{D}{b})|_{r_0}=\frac{\Phi^2}{8\pi b\ell}\big\{\zeta +3\zeta^{-1} -2\eta\zeta^{-1}(\zeta +\zeta^{-1}) \big\}.
\end{equation} The derivation of this equation is made easier by
using the relations 
$$b= 2\pi \eta\zeta\ell,\ \frac{\mathrm{d}\zeta}{\mathrm{d}\eta} =-D\frac{\zeta}{\eta},\ \frac{\mathrm{d}(\eta\zeta)}{\mathrm{d}\eta} = \eta\zeta (\zeta + \zeta^{-1}), \ \frac{\mathrm{d}D}{\mathrm{d}\eta} = -(\zeta + \zeta^{-1})+(\zeta - \zeta^{-1})D.$$
Notice that $b$ is not a constant in the energy minimization; $\Avec_o$ is not fixed by $\Phi$ alone. The subscript $\ell$ in $\Ecal_\ell$ indicates that $\ell$ is kept constant in the energy minimization.

 \begin{figure}[h] 
   \centering
   \includegraphics[width=3.5in]{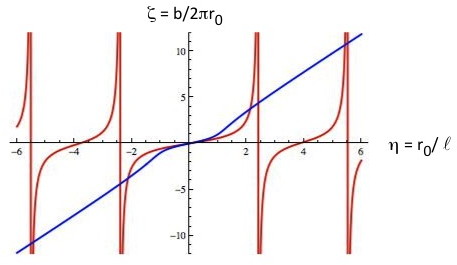} 
   \caption{\scriptsize No magnetic field outside the rope; the two first solutions for $\ell>0$ and $\ell<0$, corresponding respectively to $b>0$ (right-handed magnetic field lines of force) and $b<0$ (left-handed lines of force). The critical points are the intersections of the red curves (asymptotes excluded) and the blue curves. Red curve: $\zeta = J_1(\eta)/J_0(\eta)$ (the boundary condition); blue curve: $\partial\Ecal/\partial r_0 = 0$. See Table~\ref{t1} for numerical values.(adapted from \cite{kleman14})}
   \label{f1}
\end{figure}

The critical points are given by the intersections of
\begin{equation}
\label{e37b}
{\eta=\frac{\zeta}{2}+ \frac{\zeta}{1+\zeta^{2}}},
\end{equation} and $\zeta= {J_1(\eta)}/{J_0(\eta)}$ (the boundary conditions), see Fig.~\ref{f1}. The line energy is
 \begin{equation}
\label{e38b}
\Ecal_\ell|_{r_0} = \frac{\Phi^2}{({2\pi}{r_0})^2}\frac{3+ \zeta^2}{16}=\frac{B^2b^2}{16\pi^2}\frac{(1+\zeta^2)^2}{\zeta^2(3+ \zeta^2)}.
\end{equation}
\begin{table}[h]
\begin{tabular}{crcccc} 
\hline 
$\jmath$ & $\eta_\jmath=\frac{r_{0,\jmath}}{\ell_\jmath}$ &  $\zeta_j = \frac{b_\jmath}{2\pi r_{0,\jmath}}$  & $\frac{2\pi\ell_\jmath}{b_\jmath}$ & $\alpha=\frac{3+ \zeta_{\jmath}^2}{16}$&$\beta=\frac{(1+\zeta_{\jmath}^2)^2}{\zeta_{\jmath}^2(3+ \zeta_{\jmath}^2)}$ \\
\hline  
1 & 2.1288 & 3.7610 &0.1249&  1.07157  &0.945798\\
2 &5.4258  &10.6657&0.0173&  7.2973  &0.991511\\
3& 8.5950 &17.0733&0.0068&  18.4061  &0.996616\\
4 &11.7488  &23.4123&0.0036&  34.4460  &0.998189\\
5&14.8973  &29.7273&0.0023&  55.4195  &0.998874\\
{10}&30.6183&61.2039&0.00053&  234.307  &0.999733\\
{100}&313.3727&626.7421&4.0916$\times10^{-6}$&  24550.5 &0.999997\\
\hline 
\end{tabular}\vspace{5pt} \caption{\scriptsize The last column but one is the line energy, to a factor ${\Phi^2}/({2\pi}{r_0})^2$. The last column is the line energy, to a factor ${B^2b^2}/{16\pi^2}=({\Ical}/{c})^2$, which tends very quickly to a maximum, $\beta = 1$, when $\jmath$ increases.(adapted from  \cite{kleman14})} 
\label{t1} \end{table}

The critical points Fig.~\ref{f1} are the same as those obtained in \cite{kleman14} Sect. 5, which treats also of a tube embedded in a pure vector potential, but as the limit of a tube embedded in a vanishing irrotational magnetic field $-$ {\it i.e.}, the invariant is an electric current, not a flux. The energy is the same (Equation (35) in \cite{kleman14} identifies easily  with \eqref{e38b}). The critical points are stable.\bigskip

3.2$-$ \textsf{Merging and splitting of flux ropes} \smallskip

Reconnection of flux ropes are responsible for energy release in large amounts, see {\it e.g.}, \cite{priest00} for a review, \cite{linton06} for numerical calculations. From a global theoretical point of view, which is ours in this article, the merging of two ropes $\Phi_1,\ \Phi_2$ should yield, if it occurs, a rope $\Phi = \Phi_1+ \Phi_2$, independently of the mechanism by which this merging occurs; reciprocally a rope $\Phi$ could split, if the event is energetically favoured, into $\Phi_i$ ropes, with $\Phi=\sum_i\Phi_i$. Notice that the invariance of the current $\Ical_i$ (equivalently, of $\ell$), employed to calculate the stratum quantization, does not mean that the currents obey a conservation law of the same type: the currents are not topological invariants. 

However there is another conservation law, namely
 the conservation of the {\sf total magnetic helicity} [$\Hcal_{tot}= \iiint_V\Avec\cdot\Bvec\mathrm dV$] \cite{berger84}, which is often invoked; This law reads $$\Hcal_{tot} =\sum_i \Hcal_{i,tot},\qquad\Hcal_{i,tot}=\iiint_{V_i}\Avec\cdot\Bvec\mathrm dV,$$ where each volume of integration is bordered by surfaces to which the magnetic field is tangent, $\Bvec\cdot\nvec = 0$ (magnetic surfaces). 
 
 This law cannot be applied under this form to the merging (or splitting) of flux tubes, because such a process cannot propagate instantaneously along the full length of the tubes, even if the tubes are parallel all along; thus the initial domain of merging (or splitting) necessarily involves a finite volume of tube which is {\it not} bound by a magnetic surface. 
However, as claimed in \cite{berger84}, the conservation law still holds  
  but for another form of magnetic helicity. In the straight tube case, this {\it restricted helicity} takes the form $\Hcal_R = 4\pi\int A_\theta B_\theta r\mathrm dr$ per unit length of tube, which can be written, in the LFFF case under investigation
\
\begin{equation}
\label{e288}
\Hcal_R = 4\pi\ell^3B^2\frac{1}{J_0^2(\eta)}\int_0^\eta J_1^2(\rho) \rho\mathrm d\rho = 8\pi\ell \Ecal \frac{\zeta^2-1}{\zeta^2+1}
\end{equation}

This expression obtains from the relations already introduced, namely  
$\zeta ={J_1(\eta)}/{J_0(\eta)}$ and Eq.~\ref{e38b}, and the identity\footnote{use in sequence Eq. 11.3.31 ($\mu=-\nu = 1$) and Eq. 9.1.27 ($\nu=1$) from ref. \cite{abram}.}
$$\int_0^\eta J_1^2(\rho) \rho\mathrm d\rho =\half \eta^2\Big\{J_0^2(\eta)+J_1^2(\eta) -2\frac{J_0(\eta)J_1(\eta)}{\eta}\Big\}.$$

This conservation law is not universal; it applies to reconnection processes for tubes belonging to the same stratum. A change of stratum does not conserve magnetic helicity. More specifically we shall argue that\smallskip

i$-$ merging is favored against splitting, when the original tubes and the final one belong to the same stratum. The magnetic helicity is conserved. The amount of released energy is large, proportional to $\Ecal(n-1)/n$, where $\Ecal$ is the total energy of the tubes before their collapse to a unique one and $n$ is the number of (assumed equivalent) collapsing tubes. A change of stratum involves a different mechanism, $-$ for the sake of comparison, we make a calculation where the tubes belong to different strata, \smallskip

ii$-$ an isolated tube formerly in a stratum $\jmath > 1$ releases some energy when falling into the first stratum  $\jmath = 1$, but at the expense of topological modifications in the core, which reflects in the change of magnetic helicity.\smallskip

 a)$-$ \textsf{merging}: Consider a collection of $n$ identical parallel flux ropes, each defined by the parameters  $\Phi = Bb\ell, \ \eta,\ r_0$ and a total energy $ \Ecal \propto n \Phi^2 \alpha/r_0^2$, where $\alpha = (3+\zeta^2)/16$, (last column but one Table~\ref{t1}).\footnote{We affect these flux ropes to some stratum $\jmath=k_0$, which index we do not write explicitly.} Assume that this collection merges to a unique flux rope $\Phi_k = B_kb_k\ell_k, \ \eta_k,\ r_{0,k} , \ \Ecal _k\propto \Phi_k^2 \alpha_k/r_{0,k}^2$, where $k$ is an index $\jmath$, Table~\ref{t1}. The relations that express the conservation of the restricted magnetic helicity and of the flux are
\begin{equation}
\label{233}
\mathrm{a)}\ \ \ell\Ecal\frac{\zeta^2-1}{\zeta^2+1}
=\ell_k\Ecal_k\frac{\zeta_k^2-1}{\zeta_k^2+1},\qquad \mathrm{b)}\ \ n\Phi =\Phi_k.
\end{equation}

In the calculations that follow, we use the relations and notations
$$r_0\equiv\ell\eta, \ b\equiv 2\pi\ell\eta\zeta,\ \alpha = \frac{3+\zeta^2}{16}, \ \beta=\frac{(1+\zeta^2)^2}{\zeta^2(3+ \zeta^2)},\ u= \frac{\zeta^2-1}{\zeta^2+1}$$ where $\beta = \alpha/\eta^2$ is the energy expressed in units of $B^2b^2/(16
\pi^2)$ (last column of Table~\ref{t1}). With these modifications, the energies can be written 
$\Ecal = n\frac{\Phi^2}{\ell^2}\beta$, and $\Ecal_k = n^2\frac{\Phi^2}{\ell_k^2}\beta_k$.

Using \eqref{233}a, one gets:
\begin{equation}
\label{e311}
\frac{\Ecal}{\Ecal_k}=\frac{\ell_k}{\ell}\frac{u_k}{u}.\end{equation}Also, because $\Ecal = n\frac{\Phi^2}{\ell^2}\beta, \ \Ecal_k = n^2\frac{\Phi^2}{\ell_k^2}\beta_k$
\begin{equation}
\label{322}
\frac{\Ecal}{\Ecal_k}\equiv\frac{1}{n}(\frac{\ell_k}{\ell})^2\frac{\beta}{\beta_k}.
\end{equation} Comparing these two expressions of the ratio ${\Ecal}/{\Ecal_k}$ one gets
\begin{equation}
\label{e322}
\frac{\ell_k}{\ell} = n\frac{\beta_k}{\beta}\frac{u_k}{u},
\end{equation} and eventually
\begin{equation}
\label{e344}
\frac{\Ecal}{\Ecal_k}= n\frac{\beta_k}{\beta}\Big\{\frac{u_k}{u}\Big\}^2.
\end{equation}

If the merging of the original tubes takes place in the same stratum, this relation tells us that \begin{equation}
\label{ }
\frac{\Ecal}{\Ecal_k}= n,\qquad \triangle\Ecal=\Ecal_k -\Ecal =-\frac{n-1}{n}\Ecal.
\end{equation} Also, by using \eqref{233}b and \eqref{e322}, one gets 
\begin{equation}
\label{355}
 \frac{\ell_k}{\ell}= n, \quad \frac{B_kb_k}{Bb}= 1, \quad \frac{b_k}{b}=n,\quad \frac{r_{0,k}}{r_0}=n, \quad \frac{B}{B_k}=n.
\end{equation} \smallskip

If the original tubes and the final tube belong to different strata, we have to consider the numerical values of the quantities ${\beta_k}{u_k}^2/{\beta}{u}^2$; they all are not very different from unity, {\it e.g.,}
$${\beta_1}{u_1}^2 = 0.712495,\ {\beta_2}{u_2}^2 = 0.957251,\ {\beta_5}{u_5}^2 = 0.994362, \ {\beta_{10}}{u_{10}}^2 = 0.998666 ;$$ ${\beta_\jmath}{u_\jmath}^2 $ tends very quickly to unity, the stratum $\jmath = 1$ is strongly separated from the strata that follow. Thus the energy release does not depend much of the value of $k$ compared to $k_0$, larger than $\frac{n-1}{n}\Ecal$ if $k>k_0$, smaller otherwise. 

But there is a considerable physical difference between the merging process involving a unique stratum and the process where different strata are involved. In the first case the transition can be considered as continuous; $\eta, \ \zeta$ are constant and $r_0, \ \ell, \ B$ vary continuously. Notice that the number of pitches $r_0/\ell$ orthogonal to the tube axis does not vary. The pitches expand continuously, and the pitch $b$ along the axis expands in the same proportion. This is akin to the type of situation that has been considered up to now in numerical simulations \cite{linton06}. The merging in the discontinuous case, as we argue below, involve more drastic topological changes, and probably {\it activation processes}, possibly made favourable by the high temperature environment. \smallskip

 b)$-$ \textsf{transition to the lowest stratum}: 
Consider now an isolated tube in the $\jmath=k>1$ stratum, topological invariant $\Phi$, current intensity $\propto B_kb_k$: it can fall to the $\jmath=1$ stratum by releasing a rather small energy: the topological invariant $\Phi$ is the same (necessarily); another relation is necessary instead of the conservation of the magnetic helicity. We consider two cases. \smallskip

i)$-$ {\it conservation of the pitch orthogonal to the axis}: $\ell_1 =\ell_k$; this is equivalent to the conservation of the current: $B_1b_1=B_kb_k$. The line energies can be written:
$$\Ecal= \frac{\Phi^2}{4\pi^2 r_{0}^2} \alpha \propto \frac{\beta}{\ell^2} $$ with the appropriate indices, such that 
\begin{equation}
\label{ }
\frac{\Ecal_k}{\Ecal_1}=\frac{\beta_k}{\beta_1};
\end{equation} this small energy release, not much different from zero (see Table \ref{t1}) $\triangle \Ecal_\ell = -\Ecal_k(1-{\beta_1}/{\beta_k})$ and can be driven by quasi continuous modifications of the tube
\begin{equation}
\label{ }
\frac{\ell_{1}}{\ell_{k}}=1,\quad \frac{r_{0,1}}{r_{0,k}}=\frac{\eta_1}{\eta_k},\quad \frac{b_{1}}{b_{k}}=\frac{\zeta_1\eta_1}{\zeta_k\eta_k},\quad \frac{B_{1}}{B_{k}}=\frac{\zeta_k\eta_k}{\zeta_1\eta_1}.
\end{equation}

The numerical values in Table~\ref{t1} indicate that, while the gain in energy is rather small, the radius of the tube and the pitch along the tube axis can be significantly decreased (especially the pitch), whereas the magnetic field is concentrated, at constant flux. One can surmise that, topologically, the mechanism at the origin of the tube modification is a pinching of limited size ($\simeq b_1$) along the tube axis, which extends along the tube, accompanied by the formation of helical turns (of pitch $b_1$ along the lines of force of the magnetic field, in a coherent manner from one line to another. These modifications do not require large topological modifications; they might be continuous.

This process does not conserve the magnetic helicity. With 
$\Hcal_{R,1}= 8\pi\ell_1\Ecal_1 u_1,\ \Hcal_{R,k}= 8\pi\ell_k\Ecal_k u_k$ one gets $\triangle \Hcal_{R,\ell}=\Phi^2/2\pi\ell \Big\{\beta_1 u_1-\beta_k u_k\Big\}\neq 0$.

ii)$-$ {\it conservation of the pitch along the axis}: $b_1 = b_k$. The line energies can be written:
$$\Ecal= \frac{\Phi^2}{4\pi^2 r_{0}^2} \alpha \propto \alpha\zeta^2,$$ thus 
\begin{equation}
\label{ }
\frac{\Ecal_k}{\Ecal_1}=\frac{\alpha_k\zeta_k^2}{\alpha_1\zeta_1^2},
\end{equation}
which is considerably larger than unity; the release of energy is much larger than in the $\ell = \mathrm{constant}$ case: $\triangle\Ecal_b = \Ecal_k(1-{\alpha_1\zeta_1^2}/{\alpha_k\zeta_k^2}$), see Table~\ref{t1}; also
\begin{equation}
\label{ }
\frac{b_{1}}{b_{k}}=1,\quad \frac{r_{0,1}}{r_{0,k}}=\frac{\zeta_k}{\zeta_1},\quad \frac{\ell_{1}}{\ell_{k}}=\frac{\eta_k\zeta_k}{\eta_1\zeta_1},\quad \frac{B_{1}}{B_{k}}=\frac{\zeta_1\eta_1}{\zeta_k\eta_k}.
\end{equation}

In contrast with the previous case, the radius of the core increases and the magnetic field decreases. The transition mechanism affects now the pitch orthogonal to the axis, {\it i.e.}, the pitch of the full set of magnetic lines of force. The number of half pitches, equal to $k$ in the high energy flux tube \cite{kleman14}, is reduced to 1 in the $\jmath = 1$ configuration. This requires drastic topological changes.

This process does not conserve the magnetic helicity. One can show that $\Hcal_R = 8\pi\ell\Ecal u =\half (\Phi^2/b )\{\zeta^2-1\}$. Thus $\triangle \Hcal_{R,b}=\half\Phi^2/b \{\zeta_1^2-\zeta_k^2\}\neq 0$, a large value (compared to $\triangle \Hcal_{R,\ell}$) which is in relation with the topologically more involved transition mechanism in the $b= \mathrm{constant}$ case.
\smallskip

 c)$-$ \textsf{transition of $n$ flux tubes to the lowest stratum in two steps, at constant helicity}: In the first step, the $n$-collection merges into a unique tube belonging to the same stratum $\jmath = k_0$. Then this unique flux rope, of energy $\Ecal _{k_{0}}= \Psi^2 \alpha_{k_{0}}/r_{0,k_0}^2$, where $\Psi =n\Phi/2\pi$, falls to some other stratum $\jmath = k$, energy $\Ecal _{k}= \Psi^2 \alpha_{k}/r_{0,k}^2$, in a transition at constant magnetic helicity. Do we have a release of energy? The ratio $\Ecal _{k_{0}}/\Ecal _{k}$ can be written, applying \eqref{e344} with $n=1$
\begin{equation}
\label{}
\frac{\Ecal_{k_{0}}}{\Ecal_k}= \frac{\beta_k}{\beta_{k_{0}}}\Big\{\frac{u_k}{u_{k_{0}}}\Big\}^2.
\end{equation} A release of energy requires ${\Ecal_{k_{0}}}/{\Ecal_k}>1$, {\it i.e.,} $k>k_0$. Thus a transition to a {\it lowest} stratum with release of energy cannot be done at constant helicity. The foregoing paragraph, on the other hand, shows that  energy release is achieved for $k<k_0$ in some topological processes which do not conserve helicity, the most favorable being for $k=1$.

\section{\scriptsize{\sf THE GOLD$-$HOYLE NLFFF MODEL IN THE AB FRAMEWORK}}\label{s3}
This is a rather simple non-linear force-free model \cite{gold60},  in which field lines are uniformly twisted in the direction of the axis; it was developed with the purpose of explaining the corona flares. This model has been used, in the numerical investigation of the reconnection of tubes\cite{linton06}, or the analysis of observations of interplanetary magnetic clouds, see below. \smallskip

4.1$-$ {\sf The GH model: a reminder} \smallskip

The inside field depends on three independent parameters, the flux $\Phi$, the radius $r_0$ of the tube, and $\varpi$; the number of turns per unit length of a line of force is $\varpi/q$, where $q=4\pi^2r_0^2\varpi^2$. So each line of force has the same pitch $b=q/\varpi$; this is the main property of the GH model, which can be written $\cot\phi=2\pi\varpi r$, where $\phi$ is the angle of the lines of force with  a plane $z=$ constant. Notice that $\ell$ is the pitch along a radial line, which is a constant in the LFFF case; in the NLFFF case this is no longer true, $\ell$ is a function of $r$. Assuming $\Bvec_o=0$, in fact one has

a)$-$  \textsf{{outside}} 
\begin{equation}
\label{ }
\Avec_o =\{0, \frac{\Phi}{2\pi r}, -\Phi\varpi\},
\end{equation}

b)$-$ \textsf{{inside}} \begin{equation}
\label{ }
\Bvec =\{0,A\frac{2\pi\varpi r}{D_e}, A\frac{1}{D_e}\},\qquad \Avec_i =\{0,\frac{\Phi}{2\pi r}\frac{\ln D_e}{\ln(1+q)}, -{\Phi\varpi }\frac{\ln D_e}{\ln(1+q)}\}.
\end{equation} where
$$A =\Phi\frac{q}{\pi r_0^2 \ln(1+q)}, \qquad D_e = 1+4\pi^2\varpi^2r^2, \qquad 4\pi^2\varpi^2r_0^2 =q.$$

The energy is 
\begin{equation}
\label{eq31}
\Ecal = \half\Phi^2\varpi^2\frac{1}{\ln(1+q)}
\end{equation}

The Beltrami parameter $\alpha = 1/\ell$, which depends on $r$, is
\begin{equation}
\label{eq32}
\alpha = \frac{4\pi\varpi}{D_e}.
\end{equation}

Let us denote $$b_o= -\frac{1}{\varpi},\qquad b=\frac{q}{\varpi}, \qquad B= B_{z}(r_0)= \frac{A}{1+q},$$ $b_o$ is the pitch of the helicoids $ \psi_o =(b_o/2\pi) \theta +z$ orthogonal to the vector potential $\Avec_o$. Because $\Avec_i\cdot\Bvec_i =0$, $b=q/\varpi$ is the pitch of the lines of force of $\Bvec$ along the tube axis; it does not depend of the chosen line of force. 
One can check that $ \iint_0^{r_0} \rot \Bvec|_z r\mathrm d r\mathrm d\theta =B b$; thus the intensity takes the same expression than in the LFFF case, namely $\Ical = \frac{c}{4\pi}B b$.

Finally, if one wants, in analogy with the LFFF case, to write $\Phi=Bb\ell_{G-H}$, it is easy to check that 
$$\ell_{G-H} = \frac{(1+q)\ln(1+q)}{4\pi\varpi q}.$$ This constant is related to the Beltrami parameter $1/\alpha(r_0) = {(1+q)}/{(4\pi\varpi )}.$

The restricted magnetic helicity per unit length $\Hcal_{rel} = 4\pi\iint A_{i,\theta}B_{i,\theta} \ r \mathrm dr \mathrm d\theta$ \cite{berger84} is
\begin{equation}
\label{ }
\Hcal_{R} =\varpi \Phi^2.
\end{equation}\vspace{2pt}

4.2$-$ {\sf Stable critical points.} \smallskip

As above, we assume that the current intensity $\Ical \propto B_ib$ is a constant. This reads $\ell = constant$, {\it i.e.}, employing $\mathrm d\ell/\mathrm dq=0$,
$$\frac{\mathrm d\varpi}{\mathrm dq} \frac{(1+q)\ln(1+q)}{q\varpi}= \frac{1}{q^2}\{ q-\ln(1+q)\}.$$

The critical points are given by $\partial \Ecal/\partial q = 0$, namely
\begin{equation}
\label{31c}
\frac{\partial \Ecal}{\partial q} =  \half\Phi^{2}\frac{\varpi^2}{q(1+q)\ln^2(1+q)}\{q-2\ln(1+q)\}.
\end{equation} This expression vanishes 
for \begin{equation}
\label{32c}
q= 2.5128624\dots.
\end{equation} The second derivative, calculated at this critical point,  
$$\frac{\partial^2 \Ecal}{\partial q^2}|_{r_0}= \half\Phi^{2}\frac{\varpi^2}{q(1+q)\ln^2(1+q)}\frac{q-1}{q+1},$$
 is positive; the configuration is stable. 
Because $\cot \phi(a)= q^\half$, one gets an angle of $\phi(a)=32.2452\ ^\circ$, {independent of the radius of the tube}.

The energy at the critical point is
 \begin{equation}
\label{333}
\Ecal_{GH}|_{r_0}=\frac{\Phi^2}{({2\pi}{r_0})^2},
\end{equation} which is to be compared to the value reported in the last column but one in Table \ref{t1}. In this column the first coefficient, the smallest one, is 1.07157, which is of the same order of magnitude, and is definitely smaller than the coefficients of ${\Phi^2}/{4\pi^2 r_0^2}$ in all the other situations we have investigated, including the cases where the complementary invariant is $b$ or $B$, instead of $Bb$, see App. 2. At this point of advancement of the theory, we believe that the difference between the coefficients 1 and 1.07157 is not significant.\bigskip

4.3$-$ {\sf Merging and splitting of ropes} 

Start from a collection of $n$ flux ropes, each characterized by $\Phi, \ r_0,$ of total energy $$\Ecal = 
n\frac{\Phi^2}{(2\pi r_0)^2}\propto n \frac{q}{b^2}.$$ Their merging yields a unique flux rope $n\Phi, \ r_{0,k},$ of energy $$\Ecal_k = 
n^2\frac{\Phi^2}{(2\pi r_{0,k})^2} \propto n^2 \frac{q}{b_k^2}.$$
Thus \begin{equation}
\label{ }
\frac{\Ecal}{\Ecal_k}=\frac{1}{n}(\frac{b_k}{b})^2.
\end{equation}

The conservation of the magnetic helicity yields $b_k = n b$. Thus eventually
\begin{equation}
\label{ }
\frac{b_k}{b}=n,\qquad \frac{\Ecal}{\Ecal_k}=n.
\end{equation}

These results are similar to those obtained in the LFFF case for states all belonging to the same stratum. In the present case, there is only one stratum, very comparable to the first LFFF stratum, as indicated above. See also Eq. \eqref{355}.

\section{\sf {\scriptsize DISCUSSION}}\label{s4} 

5.1$-$ \textsf{{An extended summary of this article}}

\smallskip
 The foregoing dwells essentially upon the theoretical properties we expect for the Lundquist and Gold-Hoyle flux ropes, in the framework of a singularity model inspired in part by the theory of defects in condensed matter physics (CMP). An important concept borrowed from CMP is that of \textsf{topological invariant}, which is in our case either the current intensity $\Ical$ flowing through the tube, or the magnetic flux $\Phi$ it carries; in the first case the topological invariant is related to a singularity of an outer irrotational magnetic field $\Bvec_o$ ($\frac{4\pi}{c}\Ical=\oint_C \Bvec_o\cdot\mathrm d\svec$), in the second case to a singularity of an outer irrotational vector potential $\Avec_o$ ($\Phi=\oint_C \Avec_o\cdot\mathrm d\svec$), $C$ any loop surrounding the flux tube, the {\sf core} of the singularity in CMP parlance. The topological invariant is accompanied by another, \textit{non-topological}, invariant: $\Phi$ if the topological invariant is $\Ical$, $\Ical$ if the topological invariant is $\Phi$. This article is about the $\Phi$ topological invariant.

The field configurations of the singularity {cores} (the flux ropes themselves) are not different from those directly obtained without reference to their singular origin $-$ incidentally the flux ropes don't have to be singular, they are extended modes of the original singularities $-$, but due to this origin, they can be classified into {\sf quantized strata}. Each stratum is defined by two quantum numbers, in the LFFF case the scalars $\eta = r_0/\ell$ and $\zeta = b/(2\pi r_0)$, where $r_0$ is the radius of the rope, $\ell$ the Beltrami constant, $b$ the pitch or the period along the tube axis (the inverse of the twist); in the GH NLFFF case there is only one stratum. In each stratum the parameters defining the flux rope ($r_0, \ell, b, B$) can vary continuously, providing $\eta$ and $\zeta$ are fixed. A remarkable fact is that the first stratum $\jmath=1$ of the LFFF flux ropes (corresponding to the smallest energies) and the unique stratum of the NLFFF flux ropes are defined by
quantum numbers that are numerically very close. From the analysis in Sect. 5.2, which compares the data inferred from observations to our theory, emerges a strong feeling that all those data belong either to the LFFF first stratum, or to the unique NLFFF stratum. This is what the theory would indeed suggest. Refined measurements, in particular of $\zeta$, should mark the difference between LFFF and NLFFF.

The classification into strata plays a leading role in the merging of a collection of ropes and in the transition of a rope from one stratum to another. A well-known property used in the analysis of reconnection is that the {\sf magnetic helicity} (under its restricted form \cite{berger84}) is conserved.\footnote{We give a closed expression of the restricted helicity for a cylindrical tube, which is made possible by the quantized stratum flux rope presentation we use, Eq.~\eqref{e288}.} We  advance that this is true for merging processes taking place in the same stratum. Merging releases a large amount of energy. In a sense, this conservation law means that the merging takes place quasi continuously. On the other hand, a process that yields a transition from one stratum $\jmath =k'$ to a lower one $\jmath =k"<k'$ necessitates a change of topology $-$ but at constant topological invariant $-$ if one expects some energy release; magnetic helicity is not conserved. The relation between the magnetic topology we are alluding to and magnetic helicity, although often discussed, remains according to us an open problem.\smallskip

In the next subsection we argue that the singularity theory is justified. The last subsection compares observational data to our theoretical predictions; we believe that this comparison supports our findings.\bigskip

5.2$-$ \textsf{{Flux ropes are singularities}}\smallskip

 It is a rather remarkable fact that the observation of flux tubes indicate that the magnetic flux lines are twisted, hence their denomination of \textit{flux ropes}. Because of the boundary conditions, whatever they may be, the outside field ($\Bvec$, or $\Avec$, or both) must also be twisted and, in the assumption of irrotationality, their level sets must be helical (as Eq.~\ref{e3}) and thus behave in $1/r$, consequently singular for $r=0$. This is a very strong property of flux ropes, which is difficult to remove, because of their twisted character. \bigskip

5.3$-$ \textsf{{Comparison to observations of magnetic clouds}} \smallskip

{\it In situ} observations in the interplanetary space provide a number of data on flux ropes, including magnetic clouds. These data have been thoroughly analysed by a number of authors, in the framework of various flux rope models (Lunquist, Gold$-$Hoyle, non force-free models, $\dots$); thus we have estimates of $r_0$, $\ell$, $B_i$ and magnetic helicities in these various models, which we can compare to the present theory. Our discussion relies mostly on the data reported in ref. \cite{dasso05,lepping97,dasso03}. 

a)$-$ {\textsf {LFFF}} We pick three cases from the references above, where the raw data are analysed in the framework of the Lundquist model:

$-$ In \cite{lepping97} a flux rope observed by Wind on 18 Oct. 1995 is characterized by the values  $B= 24.3 \ \mathrm{nT}, \ r_0 = 0.135 \ \mathrm{au}, \ \alpha = 1/\ell =20 \ \mathrm{au^{-1}}$.\footnote{au = astronomical unit (approximately the length of the semi-major axis of the Earth's elliptical orbit around the sun)} These data yield $\eta = r_0/\ell \approx 2.7$.

$-$ In \cite{dasso03}, the data collected for a hot tube observed by Wind on Oct. 24-25 1995 are as follows $B = 7.2 \ \mathrm{nT}, \ r_0 = 0.035 \ \mathrm{au}, \ \alpha = 1/\ell =65.8 \ \mathrm{au^{-1}}$. These data yield $\eta \approx 2.303$.

$-$ The 'small' magnetic clouds reported in \cite{lepping97}
 yield $B$ in the range 13.8 - $15.9 \ \mathrm{nT}, \ r_0 \approx 1.6 \times 10^{-2}\ \mathrm{au}, \ \alpha = 1/\ell$ in the range 100 - $170 \ \mathrm{au^{-1}}$. Thus the value of $\eta$ stand between 1.63  and 2.72.
 
 {\sf All these values of the non-dimensional parameter $\eta$ are quite close to $\eta_1$, table 1, and in any case closer to $\eta_1$ than to $\eta_2$.} 
 This is all the more remarkable that the dispersion in the data ($r_0,\ \ell$) deduced from the observations is rather large. Notice also that there seems to be no correlation between the values of the magnetic field $B_i$ and those of $r_0$.
 
 Thus, if one adopts the value $\eta=\eta_1=2.1288$ as a best fit, then $\zeta= 3.7610$, $b=2\pi r_0$ $=0.827$ au for {\it e.g.} $r_0 = 0.035 \ \mathrm{au}$, etc.
 Let us keep to this latter case,  $r_0 = 0.035 \ \mathrm{au}$; other quantities can be calculated, as
 the \textit{energy per unit length} $\Ecal|_{r_0}$ \eqref{e38b}, which amounts to
$\Ecal|_{r_0}=0.518 \times10^4$ J/m, or the \textit{current intensity}
$\Ical=\frac{c}{4\pi}B b= 2.1323\times 10^{10}$ SI units, {\it i.e.} a current density $j_z=\frac{\Ical}{\pi r_0^2}=0.2462$ nT/s ($\pi r_0^2 = 8.66\times 10^{19}$ m$^2$). 
\smallskip

b)$-$ {\textsf {NLFFF}} Here the raw data are analysed in the framework of the Gold$-$Hoyle model \cite{dasso03} for the Wind Oct. 24-25 1995 event.  One gets $r_0 = 0.035$ au, $2\pi \varpi =46.2$ au$^{-1}$, thus $q= 2.6147$, which is remarkably close to our theoretical value $q= 2.5129$, eq.~\eqref{32c}.

This discussion can be extended to the value of the current density $j_z$. With $q=2.5129$ (the theoretical value) as above and taking for the other quantities the values from the data analysis, one gets, employing the relation $(\pi r_0^2) j_z = \frac{c}{4\pi\varpi}B_z(0) \frac{q}{1+q}$, $j_z = 0.0302$ nT/s, which is 8 times smaller than the value obtained for the same MC in the LFFF model \smallskip

{\textsf {Remark}} It is worth comparing these calculated values to the value of $j_z$ in the {\it third} model analysed in \cite{dasso03}, (a non force-free field model with constant current density) the only model that allows for an estimation of $j_z$ directly from the raw data. It appears that the favoured value is $j_z= 0.12$ nT/s, which is very different from what our NLFFF calculation above; on the other hand this estimation of $j_z$ is closer to the prediction made from the LFFF model.

\section*{\sf \scriptsize {ACKNOWLEDGMENTS}}

\scriptsize I am indebted to J. M. Robbins for long-lasting fruitful exchanges on the space physics of electromagnetism. I thank P. D\'emoulin for pointing to me the possible relevance of the flux tube singularity theory to magnetic clouds. Both made relevant remarks on the text. I am particularly grateful to J.-L. Le Mou\"el who read the manuscript with the utmost care and with whom I have had fruitful discussions.

    \newpage
\footnotesize 



\begin{thebibliography}{10}

\bibitem{kleman14}
M.~Kleman and J.~M. Robbins.
\newblock {Tubes of magnetic flux and electric current in space physics}.
\newblock {\em Solar Physics}, 289:1173--1192, 2014.
\newblock {DOI}: 10.1007/s11207-013-0399-0.

\bibitem{cantarella02}
J.~Cantarella, D.~DeTurck, and H.~Gluck.
\newblock {Vector calculus and the topology of domains in 3-space}.
\newblock {\em Amer. Math. Monthly}, 109:409--442, 2002.

\bibitem{mahajan98}
S.~M. Mahajan and Z.~Yoshida.
\newblock Double curl {B}eltrami flow: Diamagnetic structures.
\newblock {\em Phys. Rev. Lett.}, 81:4863--4866, 1998.

\bibitem{klfr08}
M.~Kleman and J.~Friedel.
\newblock Disclinations, dislocations and continuous defects: a reappraisal.
\newblock {\em Rev. Mod. Phys.}, 80:61, 2008.

\bibitem{griffiths99}
D.~J. Griffiths.
\newblock {\em {Introduction to Electrodynamics}}.
\newblock Prentice Hall, Upper Saddle River, New Jersey, 1999.
\newblock 3rd Edition.

\bibitem{dasso05}
S.~Dasso, C.H. Mandrini, P.~D\'emoulin, M.L. Luoni, and A.M. Gulisano.
\newblock Large scale {MHD} properties of interplanetary magnetic clouds.
\newblock {\em Adv. Space Res.}, 35:711--724, 2005.

\bibitem{stenflo78}
J.~O. Stenflo.
\newblock The measurement of solar magnetic fields.
\newblock {\em Rep. Progr. Phys.}, 41:865--907, 1978.

\bibitem{aharonov59}
Y.~Aharonov and D.~Bohm.
\newblock Significance of electromagnetic potentials in quantum theory.
\newblock {\em Phys. Rev.}, 115:485--491, 1959.

\bibitem{lundquist50}
S.~Lundquist.
\newblock Magnetohydrostatic fields.
\newblock {\em Arkiv Fysik}, 2:361--365, 1950.

\bibitem{gold60}
T.~Gold and F.~Hoyle.
\newblock On the origin of solar flares.
\newblock {\em MNRAS}, 120:89--105, 1960.

\bibitem{lepping97}
R.~P. Lepping, L.~F. Burlaga, A.~Szabo, K.~W. Ogilvie, W.~H. Mish,
  D.~Vassiliadis, A.~J. Lazarus, J.~T. Steinberg, C.~J. Farrugia, L.~Janoo, and
  F.~Mariani.
\newblock The {W}ind magnetic cloud and events of october 18$-$20, 1995:
  Interplanetary properties and as triggers for geomagnetic activity.
\newblock {\em J. Geophys. Res.}, 102:14049--14063, 1997.

\bibitem{dasso03}
S.~Dasso, C.~H. Mandrini, and P.~D\'emoulin.
\newblock {\em The magnetic helicity of an interplanetary hot flux tube}, pages
  786--789.
\newblock Solar Wind Ten: Proceedings of the tenth international solar wind
  conference. American Institute of Physics, CP679, 2003.
\newblock edited by M. Velli, R. Bruno and F. Malara.

\bibitem{demoulin14}
P.~D\'emoulin.
\newblock {\em Evolution of interplanetary coronal mass ejections and magnetic
  clouds in the heliosphere}, pages 1--10.
\newblock Nature of prominences and their role in Space Weather. IAU Symposium,
  volume S300, 2014.
\newblock edited by B. Schmieder, J. M. Malherbe and S. T. Wu.

\bibitem{abram}
M.~Abramowicz and I.~A. Stegun, editors.
\newblock {\em Handbook of Mathematical Functions}.
\newblock Dover, New York, 1970.

\bibitem{priest00}
E.~Priest and T.~Forbes.
\newblock {\em Magnetic Reconnection: \footnotesize{MHD theory and
  applications}}.
\newblock Cambridge University Press, UK, 2000.

\bibitem{linton06}
M.~G. Linton.
\newblock Reconnection of nonidentical flux tubes.
\newblock {\em J. Geophys. Res.}, 111:A12S09, 2006.

\bibitem{berger84}
M.~A. Berger and G.~B. Field.
\newblock The topological properties of magnetic helicity.
\newblock {\em J. Fluid Mech.}, 147:133--148, 1984.

\end{thebibliography}
\end{document}